\documentclass[prb,aps,twocolumn]{revtex4}
\usepackage{graphicx}

\newcommand{\beq}{\begin{equation}}
\newcommand{\eeq}{\end{equation}}

\newcommand{\bea}{\begin{eqnarray}}
\newcommand{\eea}{\end{eqnarray}}
\newcommand{\etal}{{\em et al.}}

\def\jour#1#2#3#4{{#1}{\bf #2}, #3 (#4)}
\def\tit#1#2#3#4#5{{#1}{\bf #2}, #3 (#4)}

\def\rmp{Rev.\ Mod.\ Phys.\ }

\def\pr{Phys.\ Rev.\ }
\def\prb{Phys.\ Rev.\ B\ }

\def\jpa{J.\ Phys.\ A\ }

\def\natu{Nature\ }

\def\vecb#1{{\mathbf #1}}

\begin{document}

\title{Interplay of quantum and thermal fluctuations in a frustrated magnet}

\author{S. V. Isakov$^1$ and R. Moessner$^2$}

\affiliation{$^1$Department of Physics, Stockholm University, AlbaNova,
S-106 91 Stockholm, Sweden}
\affiliation{$^2$Laboratoire de Physique Th\'eorique de l'Ecole Normale
Sup\'erieure, CNRS-UMR8549, Paris, France}

\begin{abstract}
We demonstrate the presence of an extended critical phase in the
transverse field Ising magnet on the triangular lattice, in a regime
where both thermal and quantum fluctuations are important. We map out
a complete phase diagram by means of quantum Monte Carlo simulations,
and find that the critical phase is the result of thermal
fluctuations destabilising an order established by the quantum
fluctuations. It is separated by two Kosterlitz-Thouless transitions
from the paramagnet on one hand and the quantum-fluctuation driven
three-sublattice ordered phase on the other. Our work provides further
evidence that the zero temperature quantum phase transition is in the
3d $XY$ universality class.
\end{abstract}

\maketitle

\section{Introduction}
In geometrically frustrated magnets, the arrangement of the spins on a
lattice precludes the establishment of a simple collinear N\'eel
state. In the absence of such an `obvious' ordering pattern, these
magnets are often subject to a large number of competing
instabilities. This endows them with a rich behaviour, manifested in
unconventional phases, phase transitions and excitations, a large
low-energy density of states and unusual dynamical
properties.\cite{reviews}

When added to a frustrated magnet which is not ordered classically at
temperature $T=0$, both thermal and quantum fluctuations often
generate an ordering transition, a process known as order by
disorder.\cite{villainobdo,shenderobdo,henleyobdo} Whereas quantum
order by disorder occurs in a fashion quite analogous to the ordering
induced by thermal fluctuations, the nature of the ordering (if any)
of even some `simple' model systems (such as the Heisenberg magnet on
the pyrochlore lattice) in the presence of stronger quantum
fluctuations remains unclear.

The interplay of thermal and quantum fluctuations has so far not
received a great deal of theoretical attention for frustrated
magnets. By contrast, there has been a considerable amount of interest
devoted to the real-time dynamics and transport at finite temperature
near quantum critical points.\cite{kedar} In addition, this
combination has even been claimed to provide an alternative use of
quantum effects for speeding up computations.\cite{appros}

In this publication, we study the combined effect of thermal and
quantum fluctuations on possibly the simplest quantum frustrated
magnet, namely an Ising model on the triangular lattice in a
transverse magnetic field. The manifold of classical ground states is
hugely degenerate, and correlations averaged over it are
critical.\cite{wannhout} This model thus enables us to investigate how
the two types of fluctuations together establish and destroy order
out of the exponentially large set of classical ground states. By
changing the values of transverse field, $\Gamma$, and temperature,
$T$, their strengths are in principle separately experimentally
tunable.

We map out the phase diagram of the Ising model on the triangular
lattice in the transverse field-temperature plane. The bulk of the
paper is devoted to a numerical study based on a continuous time
quantum Monte Carlo algorithm. We find that there are three different
phases. In the regime where quantum fluctuations are weak, yet
dominate over thermal fluctuations, they generate a three-sublattice
ordered phase. Upon increasing the strength of thermal fluctuations,
it `melts' into a critical phase which has a finite extent and
displays drifting exponents. This dome-shaped critical phase is
delineated above and below by Kosterlitz-Thouless (KT) transitions at
finite temperature. At $T=0$, it terminates in the classical critical
point at zero field and in a quantum phase transition at a finite
$\Gamma_c$. At high temperatures and large field strengths, one
retrieves a simple paramagnet -- order by disorder is a reentrant
phenomenon. We briefly comment on the advantages of different
diagnostics for the floating KT phase.

The results reported here largely build on the following pieces of
work. In Ref.~\onlinecite{ftfim}, a number of frustrated transverse
field Ising models were studied, and a connection was pointed out
between these models and classical stacked magnets in a scaling
limit. An influential paper by Blankschtein and
coworkers\cite{blankschtein} on stacked triangular magnets derived a
Landau-Ginzburg-Wilson theory for these systems, and predicted the
stacked problem to be in the universality class of the $XY$ model in
three dimensions with a six-state clock term, a problem which has in
turn been studied by Jose \etal\cite{Jose77} Collecting these results
together led to a conjecture of the phase diagram we map out in the
following.\cite{ftfim} We note that, by adding a continuous
degree of freedom to a classical triangular Ising
antiferromagnet,
a floating critical phase has
previously been obtained by Chandra \etal\cite{premi94}

The general validity of the Landau-Ginzburg approach has been somewhat
called into question by Monte Carlo simulations on the
ferromagnetically stacked triangular Ising antiferromagnet, with
several groups presenting evidence in favour of critical behaviour in
the three-dimensional universality class, while others found a
different, possibly new, universality class to be a more likely
scenario.\cite{heino89,tricrit,bunker,plumer95,collinsrev} 
Our work here lends
further support to the Landau-Ginzburg approach which is independent
of the previous numerical simulations.

\section{Model}
Possibly the simplest model which has both a high classical ground
state degeneracy and a non-trivial quantum dynamics is the transverse
magnetic field antiferromagnetic Ising model on the triangular
lattice. The model has the following Hamiltonian:
\beq H = J\sum_{\langle i,j \rangle}
	\sigma^{z}_{i} \sigma^{z}_{j} - \Gamma \sum_{i} \sigma^{x}_{i},
		\label{qu_hamiltonian}
\eeq
where $J>0$ is nearest-neighbour exchange couplings and $\Gamma$ is the
magnetic field strength. 

The classical model triangular Ising magnet is obtained in the absence
of a transverse field, $\Gamma=0$. Its solution is well
known:\cite{wannhout} it is disordered at any finite temperature, and
critical at $T=0$, where it retains an extensive zero point entropy,
${\cal S}$, with ${\cal S}/N=0.323 k_B$, where $N$ is the number of
spins and $k_B$ denotes the Boltzmann constant. Upon application of an
infinitesimal transverse field, $\Gamma=0^+$, at $T=0$, the magnet
orders into a three sublattice $\sqrt{3}\times\sqrt{3}$ pattern with a
sublattice magnetisation pattern $(1,0,-1)$.\cite{ftfim}

\subsection{Mapping to a stacked magnet}
Using the Suzuki-Trotter formalism,\cite{Suzuki71,Trotter59} one can
map the d-dimensional quantum model onto the (d+1)-dimensional
classical problem. The partition function of the quantum Hamiltonian
(Eq.~\ref{qu_hamiltonian}), ${\rm Tr} \exp(-\beta H)$, is equivalent to
the following partition function:
\bea
	& Z = {\rm Tr} \exp(-S_{cl}), & \nonumber \\
	& S_{cl} = K \sum_{\langle i,j \rangle,\tau} S_{i,\tau}S_{j,\tau}-K^{\tau}
		\sum_{i,\tau} S_{i,\tau}S_{i,\tau+1}, &
	\label{cl_hamiltonian}
\eea
where $S_{i,\tau}$ are classical Ising spins taking values $1$ or $-1$;
$\tau$ is the index in the imaginary time direction and runs from $0$ to $N$,
$N=\beta/\epsilon$ is the number of slices in the imaginary time direction and
$\epsilon$ is the imaginary time step; $K=J\epsilon$ and
$K^{\tau} = -\frac{1}{2}\ln\tanh\epsilon\Gamma$. The mapping becomes exact in
the scaling limit $\epsilon\rightarrow0, K^{\tau}\rightarrow\infty$, while
maintaining 
\beq
	e^{-2K^{\tau}} = \epsilon\Gamma\ .
	\label{scaling_limit}
\eeq
It is in this formulation that the dimensional crossover between the
high temperature $d=2$ and the low temperature $d=3$ behaviour is most
transparent. Besides providing an efficient Monte Carlo algorithm (see
below), the above mapping also allows us to make contact with the
literature on stacked triangular magnets,\cite{collinsrev} from which
our problem can be obtained in the scaling analysis of infinite
coupling strength in the third dimension, with the third dimension
being of finite length at nonzero temperature.

\subsection{LGW analysis}
Blankschtein \etal\cite{blankschtein} have identified a complex XY
order parameter for a stacked triangular antiferromagnet: the amplitude
and phase of the Fourier mode at $\vec{Q}_\pm=(\pm4\pi/3,0)$,
$\psi_{\pm}=m \exp(\pm i\phi)$. They have obtained the following LGW
Hamiltonian (see below):
\bea
	H_{LGW} = \sum_{\vec{q}} (r+q^{2})m^{2}+u_{4}m^{4} \nonumber \\
	+u_{6}m^{6}+v_{6}m^6\cos(6\theta).
\eea
A six-fold clock anisotropy thus appears at sixth order in $m$.  The
six-fold clock term is irrelevant at $d=3$ so that the transition into
a phase with nonzero $m$ should be in the $d=3$ XY universality
class. This applies both to the zero temperature quantum phase
transition as we vary $\Gamma$ and to the corresponding transition in
a stacked Ising magnet. With the clock term being dangerously
irrelevant, the transition is immediately into a state with a broken
clock symmetry, the details of which depend on the sign of $v_6$ (see
below).

At finite temperatures the triangular transverse field Ising model
maps onto a stacked magnet of finite size, $L_\tau$, in the temporal
direction, which is effectively $2$-dimensional as the temporal
correlation length, $\xi_\tau$, exceeds $L_\tau$ close to a continuous
phase transition. The $2$-dimensional six-state clock model, studied
by Jose \etal,\cite{Jose77} has a remarkable phase diagram consisting
of three phases: a disordered phase, an extended intermediate KT
phase, and an ordered phase. The extended KT phase owes is existence
to the fact that in $d=2$, the clock term does not become relevant
immediately below the KT transition of the XY model; instead, the
coupling needs to be increased further until the clock term is
sufficiently strong to assert itself.

One of the central objectives of this paper is to establish the
presence of this floating KT phase as a finite-temperature induced
dimensional crossover in the $2+1$ dimensional quantum Ising magnet on
the triangular lattice. In our context, the presence of such a phase
would be the result of an XY order parameter generated by frustration
and stabilised by quantum fluctuations, together with destabilising
fluctuations provided by the finite temperature.

Details of the ordered state depend on the sign of $v_{6}$. $H_{LGW}$
is minimised by $M=|M|\exp(i\Phi)$, with
$\Phi=n\pi/3$ and
$\Phi=(n+1/2)\pi/3$ for $v_{6}<0$ and
$v_{6}>0$ respectively, with $n=1\ldots6$. 
The real space configurations of the ordered
phases can be obtained by Fourier transforming these modes. In the
case of $v_{6}<0$, Fourier transforming yields the following
magnetisations assigned to the three sublattices of the triangular
lattice: $|M|(1,-1/2,-1/2)$.  We depict this phase as
$(+--)$. In the case of $v_{6}>0$, we have the
following sublattice magnetisations:
$|M|(\sqrt{3}/2,0,-\sqrt{3}/2)$. We depict this phase as
$(+0-)$. There are six degenerate states for both $v_{6}<0$ and
$v_{6}>0$.

It is difficult to determine the sign of $v_{6}$ -- in particular, in
the effective Hamiltonian, its value may drift and even change
sign.\cite{blankschtein,miyashita} The partition function
$Z=\exp(-\sum_{ij}s_{i}K_{ij}s_{j})$ of the Ising model $H=(1/\beta)
\sum_{ij} s_{i}K_{ij}s_{j}$ is equivalent to the following partition
function (see, e.g. Ref.~\onlinecite{amit})
\bea
	Z\propto \int D \psi \exp \left( -\cal{L} \right)=
	\int D \psi \exp \biggl\{ -\sum_{ij} \psi_{i}K_{ij}\psi_{j} \nonumber \\
		+\sum_{i}\ln\cosh \Bigl( \sum_{j} 2K_{ij}\psi_{j} \Bigr) \biggr\},
\eea

The first term is the interaction matrix, which determines the soft modes
$\psi_\pm$ to be
located at $Q_\pm$ (Eq.~\ref{int_matrix}).
Performing the Fourier transform and expanding
$\ln\cosh(x)=\frac{x^{2}}{2}-\frac{x^{4}}{12}+\frac{x^{6}}{45}
	-\frac{17x^{8}}{2520}+\ldots$, we obtain 
the following expression for the contribution of the non-linear term
to
$\cal{L}$
\bea
{\cal{L}}_{nl}&=&
	-2\sum_{\vecb{k}}K(\vecb{k})K(-\vecb{k})\psi(\vecb{k})
		\psi(-\vecb{k}) \nonumber \\
	&+&\frac{4}{3} \sum_{\vecb{k}_{i}}\left[\prod_{i=1}^4\left(
		K(\vecb{k_{i}})\psi(\vecb{k}_{i})\right)
	\delta^\prime(\vecb{k}_{1}+\ldots+\vecb{k}_{4})\right] \\
	&-&\frac{64}{45} \sum_{\vecb{k}_{i}}\left[\prod_{i=1}^6\left(
		K(\vecb{k_{i}})\psi(\vecb{k}_{i})\right)
	\delta^\prime(\vecb{k}_{1}+\ldots+\vecb{k}_{6})\right]\nonumber
\ldots.
	\label{lagr}
\eea

The interactions for the stacked triangular antiferromagnet are
written in Fourier representation as
\bea
	K(\vecb{k})=\sum_{k_{x},k_{y},k_{z}} \left( J\left[ \cos(k_{x}) +
	\cos(k_{x}/2+\sqrt{3}k_{y}/2) \right. \right.\nonumber \\
	\left.	\left.
	+\cos(k_{x}/2-\sqrt{3}k_{y}/2)\right]
	-J'\cos(k_{z}) \right), 
	\label{int_matrix}
\eea
where the sum is over the hexagonal Brillouin zone. The LGW
Hamiltonian is thus constructed in terms of the coefficients
$m\exp(\pm i \phi)$), varying slowly in space, of the `soft' modes
$\psi_\pm$.

The primed delta functions indicate that the wavevectors
$\vecb{k_{i}}$ need to add up to a reciprocal lattice vector. This
requirement leads to the XY nature of the effective theory at low
order.
At sixth order, $\psi^{3}_{+}\psi^{3}_{-}$ and
$\psi^{6}_{+}+\psi^{6}_{-}$ occur, the latter being the clock term, $m
\cos(6\theta)$. $v_{6}<0$ at this order. However, the $\cos(6\theta)$
term appears at higher orders as well, e.g.,
$\psi_{+}\psi_{-}(\psi^{6}_{+}+\psi^{6}_{-})$ at eighth order. Since
the terms in the series have alternating sign and large coefficients,
one cannot reliably determine the sign of $v_6$ near the transition
this way.

\section{Monte Carlo method}
\subsection{The quantum Monte Carlo algorithm}
The $d=2+1$ dimensional classical problem obtained this way has only
positive weights -- there is no `sign problem'.
It can therefore be studied reasonably straightforwardly by 
Monte Carlo simulations. We thus 
simulate the classical problem
defined by Eq.~\ref{cl_hamiltonian}. 
However, 
it is difficult to simulate the discretized
version of Eq.~\ref{cl_hamiltonian} because of the scaling limit
(Eq.~\ref{scaling_limit}):
in order to avoid discretisation
errors, one has to take a very large ferromagnetic coupling in the
imaginary time direction $K^{\tau}\rightarrow\infty$ 
at the same time increasing the
height of the system exponentially as $e^{2K^{\tau}}$.

In order to avoid this problem, we use a continuous time
algorithm.\cite{beardwiese, prokofev, Rieger99} The basic idea behind this algorithm is
that, in the scaling limit, the density of domain walls in the
imaginary time direction becomes exponentially sparse and it is thus
more efficient to keep track of the location of the domain walls,
using $\exp(2K^\tau)$ as a unit of length. Thus, the height of the
system, $N/\exp(2K^\tau)=\beta/(\epsilon\exp(2K^\tau))=\beta\Gamma$,
measured in units of $\exp(2K^\tau)$, remains fixed in the continuum
limit.

Due to the frustrated nature of our problem we cannot use a cluster
algorithm\cite{Swendsen87,Wolff89,Krauthreview} in the space
directions as one could for the case of unfrustrated magnets; however,
we can use a cluster algorithm in the time direction. The algorithm
works as follows (for more details, see
Ref.~\onlinecite{Rieger99}). We pick a random site on the triangular
$L\times L$ lattice and build a cluster on that site in the imaginary
time direction. Its length, $\tau$, is distributed according to the
probability distribution $P(\tau)\propto\exp(-\Gamma\tau)$ -- this
prescription eliminates a `freezing' of the algorithm due to the
diverging temporal coupling.  A given cluster is flipped using the
Metropolis prescription in the spatial direction, {\em i.e.} we flip
the cluster with probability $p=\min(1,
\exp(-\Delta E))$, where $\Delta E $ is the (spatial) 
energy difference between
the original configuration and the configuration with a flipped
cluster.

One check of our Monte Carlo algorithm consisted of comparing it with
the diagonalisation of a $3\times3$ lattice, and we have found
excellent agreement between the two approaches. 

\subsection{Parameters of the simulations}

We impose periodic boundary condition on the triangular lattice and
performed simulations on lattices of size
$L=9,12,15,18,24,30,36,48,60,75$.  We have estimated the correlation
time $\tau_{c}$ for different parameters. We perform usually
$500\tau_{c}$ Monte Carlo steps (MCS) for equilibration and from
$10^4\tau_{c}$ to $5\cdot10^5\tau_{c}$ MCS for averaging.

\section{Scaling analysis}

In this section, we outline the quantities useful for detecting a
possible floating KT phase and the adjacent KT phase transitions as
well as the ordered phase.

\subsection{Order parameters}
The complex XY order parameter deduced from the LGW analysis is given by:
\beq
	m e^{i \theta} \equiv (m_{1} + m_{2} e^{i \frac{4\pi}{3}} + 
m_{3} e^{i \frac{-4\pi}{3}})/\sqrt{3} 
	\label{order_param}
\eeq
where the $m_{i}$ are the magnetisations of the
three sublattices, and  $m$ is real and positive.
$m$ is close to 1 in the limit of zero temperature and vanishing
transverse field. It equals zero for the disordered and KT phases, 
vanishing exponentially and 
algebraically in the limit the system size
$L\rightarrow\infty$. 
There, the corresponding susceptibility is
\beq
	\chi = L^{2} \langle m^{2} \rangle / T.
\label{sus}
\eeq
To detect clock symmetry breaking,
we consider
\beq
	c_{6} = \frac{\langle m^{6} \cos(6\theta) \rangle}{\langle m^{6} \rangle}.
\label{cos6theta}
\eeq
It is easy to check that $c_{6}$ equals zero for disordered and KT
phases, $c_{6}$ equals $-1$ for the $(+0-)$ phase, and $c_{6}$ equals
$1$ for the $(+--)$ phase. We have chosen not directly to average
$\cos(6 \theta)$ as its value fluctuates most strongly when $m$ is
small, that is to say, the ordering we are trying to determine the
details of is weakest. The exponent $m^6$ has been chosen, somewhat
arbitrarily, as the corresponding term in the LGW action is
$\psi_+^6+\psi_-^6=m^6 \cos(6\theta)$.

\subsection{Binder cumulant}
The appropriate Binder cumulant is:\cite{binder}
\beq
	U = 1 - \frac{\langle m^{4} \rangle}{3 \langle m^{2} \rangle^{2}}.
\eeq
The Binder cumulant has a scaling dimension of zero. It thus has the
advantage of not requiring fitting an unknown leading exponent. In the
limit $L\rightarrow\infty$, the Binder cumulant has the following
behaviour: $U_{L}\rightarrow 0$ at disordered phase, $U_{L}\rightarrow
2/3$ at ordered phase, and $U_{L}\rightarrow U^{*}$ at a critical
point.  For an extended critical phase, the value of $U^{*}$ can
drift, so that its value depends on the precise location within the
critical phase.

Since the Binder cumulant has a scaling dimension of zero, the curves
for different system sizes at a critical coupling should fall on a
line of points where $U_{L}=U_{L'}$ (at least in the region where
corrections to scaling are small). As we describe below, this
criterion in fact provides an overestimate of the size of the critical
phase.

\subsection{Locating the KT transitions}
Here, we briefly describe the finite-size scaling analysis appropriate
for KT transitions.  This analysis follows that used by Challa and
Landau.\cite{challalandau} In order to determine the presence of a KT
phase, we check if we can fit our simulations to the scaling
forms predicted by KT theory. \cite{Kosterlitz74} This has the
advantage of restricting the number of fitting parameters, which in
the most general case (allowing for critical exponents different from
the KT ones) would be too large to be practical.

In particular, the correlation length $\xi$ and the susceptibility
$\chi$ behave as
\bea
	\xi \propto \exp(at^{-1/2}), \nonumber \\
	\chi \propto \xi^{2-\eta}, \nonumber \\
	m \propto \xi^{-\eta/2}
	\label{m_chi_kt}
\eea
where $a$ is a nonuniversal constant, $t=(T-T_{c})/T_{c}$ is a reduced
temperature, and $T_{c}$ is a critical temperature.

The finite-size scaling form of order parameter and susceptibility is
given by 
\bea
	m_{L} = L^{-b} m_{0}(\xi/L), \nonumber \\
	\chi_{L} = L^{c} \chi_{0}(\xi/L),
	\label{m_chi_kt2}
\eea
where $m_{0}$ and $\chi_{0}$ are unknown universal functions and $b$ and $c$
are constants. It follows from (\ref{m_chi_kt}) that in the infinite system
size limit, we have $b=\eta/2$ and $c=2-\eta$. Therefore, at a
critical point, one has:
\bea
	m_{L} \propto L^{-\eta/2},\\
	\label{m_l}
	\chi_{L} \propto L^{2-\eta}.
\eea
If we have an extended critical phase, these relations should hold over
finite temperature range from the upper critical temperature $T_{2}$ to
the lower critical temperature $T_{1}$. The critical exponent $\eta$ should
vary continuously from $T_{2}$ to $T_{1}$. Plotting $\ln(m_{L})$ or
$\ln(\chi_{L})$ versus $\ln(L)$, we can find the critical exponent $\eta$ at
any point of the extended critical phase.

We can rewrite Eq.~(\ref{m_chi_kt2}) as
\bea
	m_{L} L^{b} = m_{0}(L^{-1}\exp(at^{-1/2})),
	\label{m_kt} \\
	\chi_{L} L^{-c} = \chi_{0}(L^{-1}\exp(at^{-1/2})).
	\label{chi_kt}
\eea
Eq.~(\ref{m_kt}) is valid for $T<T_{1}$ and (\ref{chi_kt}) is valid
for $T>T_{2}$.  For an appropriately chosen set of parameters $a, c,
T_{1}$ the plot of $\chi_{L} L^{-c}$ versus $L^{-1}\exp(at^{-1/2})$
should collapse onto a universal curve for different system sizes
$L$. The same should hold for a plot of $m_{L} L^{b}$ versus
$L^{-1}\exp(at^{-1/2})$. From such a fit, the upper and lower critical
temperatures can be determined.

\section{Results}
\begin{figure}[ht] 
\centerline{\includegraphics[angle=-90, width=3in]{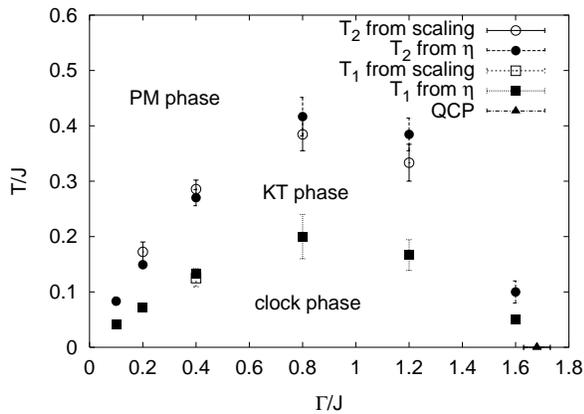}}
\caption{The phase diagram of the transverse field Ising model
on the triangular lattice. A floating KT phase separates the
ordered clock phase from the paramagnet (PM). QCP denotes the location
of the quantum critical point.}
\label{fig:phase_diagram}
\end{figure}

In Fig.~\ref{fig:phase_diagram} we present the phase diagram that we find
from Monte Carlo simulations. There are three phases: a disordered phase at
high temperatures ($T\gg J$) 
or large magnetic field strengths $\Gamma\gg J$, 
an extended
KT phase at intermediate temperatures, and an $(+0-)$ ordered phase at
low temperatures.

\subsection{The KT phase at $\Gamma/J=0.4$}

Our most complete data set was taken for $\Gamma/J=0.4$, for which we
describe our analysis in detail in the next few paragraphs. We have
chosen this value of $\Gamma$ as a compromise between
the following
requirements. Firstly, we want to stay well clear of the
zero temperature 
critical points at $\Gamma=0$ and larger $\Gamma=\Gamma_c$ 
and concomitant
possible crossover phenomena. Secondly, a high transition temperature is
needed to allow us to simulate systems with only moderate extent in the
imaginary time direction. Thirdly, as our algorithm slows down as the
density of domain walls in the imaginary time direction increases, we
would like to choose weak quantum fluctuations, that is to say, small
values of $\Gamma$. $\Gamma/J=0.4$, located left of centre of
the KT dome (Fig.~\ref{fig:phase_diagram})
thus appears to be a sensible choice.

\begin{figure}[ht]
{\centerline{\includegraphics[angle=-90, width=3in]{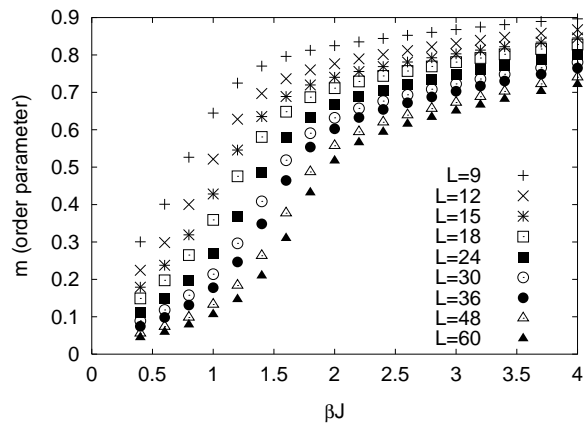}}
\caption{The order parameter $m$ versus $\beta J$ for different
system sizes.}
\label{fig:op_beta}}
\end{figure}

First, Fig.~\ref{fig:op_beta} shows the behaviour of the order
parameter $m$ as a function of inverse temperature $\beta$ for
different system sizes. One can notice easily that the order parameter
has strong dependence on the system size in wide range of temperatures
-- it decreases with increasing system size.

\begin{figure}[ht]
{\centerline{\includegraphics[angle=-90, width=3in]{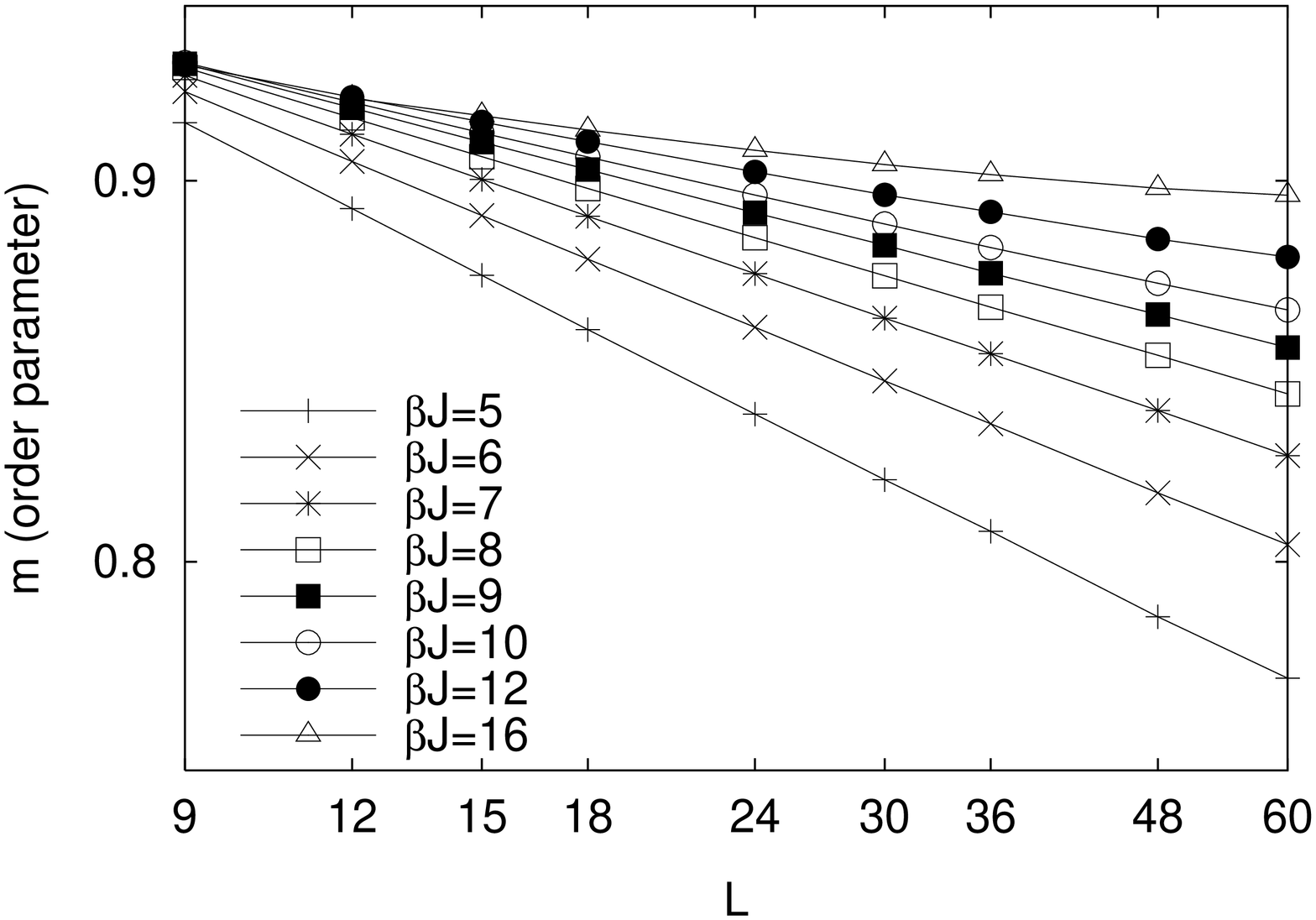}}
\centerline{\includegraphics[angle=-90, width=3in]{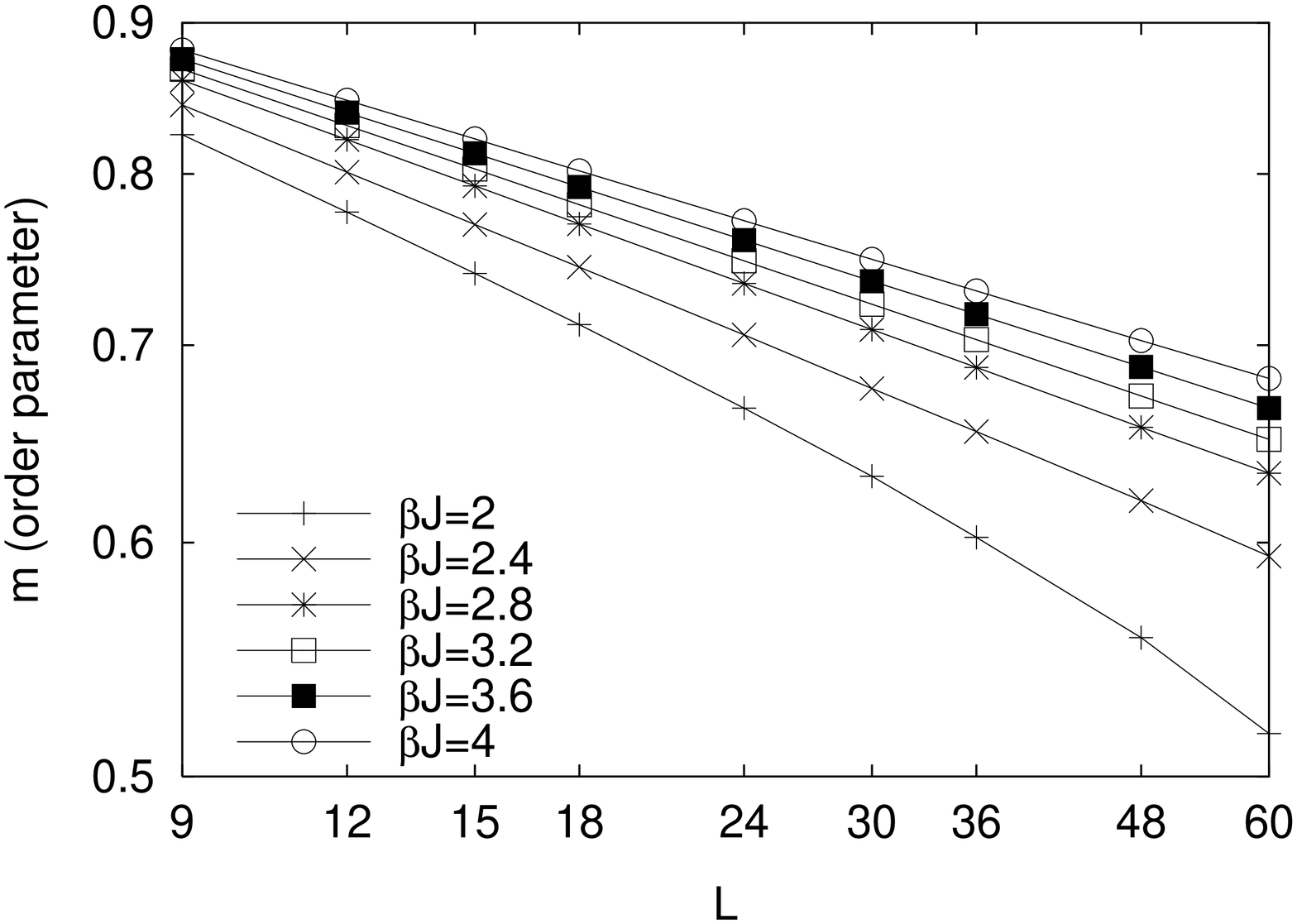}}
\caption{The order parameter $m$ versus system size $L$ at various
temperatures. The lines connect the points to guide the eye.}
\label{fig:op_l}}
\end{figure}

Fig.~\ref{fig:op_l} is a doubly logarithmic plot of the behaviour of
the order parameter as a function of the system size.  At high
temperatures, it extrapolates to zero, whereas it levels out to a
constant at low temperatures. For a wide range of temperatures in
between, the data points display linear behaviour (without any
evidence of
logarithmic corrections), beginning at $\beta
J=2.8$ down to $\beta J=9.0$. The exponent $\eta$ varies continuously
from $0.323$ at $\beta J=2.8$ to $0.092$ at $\beta J=9.0$. 

This is indeed close to the range expected from KT finite-size scaling
(\ref{m_l}), although there is an overestimate of the size of the KT
phase when compared to the expected range of critical exponents, which
is between 1/4 and 1/9.  As in the case of the Binder cumulant below,
this is a simple consequence of the fact that the difference between
the ordered and the critical phase is indiscernible when the size of
the system is much less than the correlation length. As the latter
grows near the transition, increasingly large finite-size systems in
the ordered phase appear to be critical.

\begin{figure}[ht]
{\centerline{\includegraphics[angle=-90, width=3in]{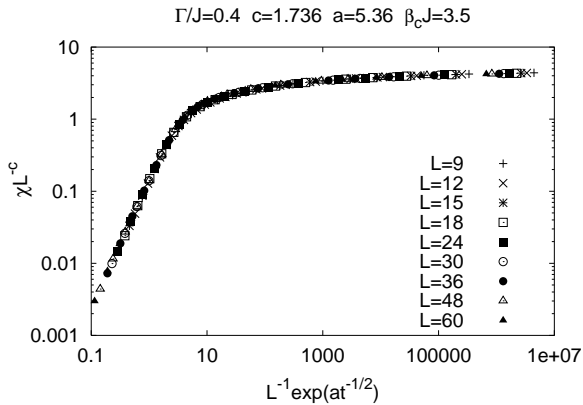}}
\caption{Data collapse of the susceptibility for the upper transition.}
\label{fig:ht_fss}}
\end{figure}

To check whether this critical region is indeed delineated by KT
transitions, in Fig.~\ref{fig:ht_fss} we show the data collapse
(Eq.~\ref{chi_kt}) for the upper transition. The data scales very well
with the following set of parameters $\beta_{2} J=3.5, a=5.36,
c=1.736$. Thus we can conclude that the system has a transition between
the disordered phase and KT phase at $\beta_{2} J=3.5\pm0.2$. The
critical exponent $\eta=0.263\pm0.015$ at the transition point. This
value of the critical exponent $\eta$ is close to the theoretical
prediction $1/4$, and the overestimate appears to be part of a
systematic trend discussed below.

\begin{figure}[ht]
{\centerline{\includegraphics[angle=-90, width=3in]{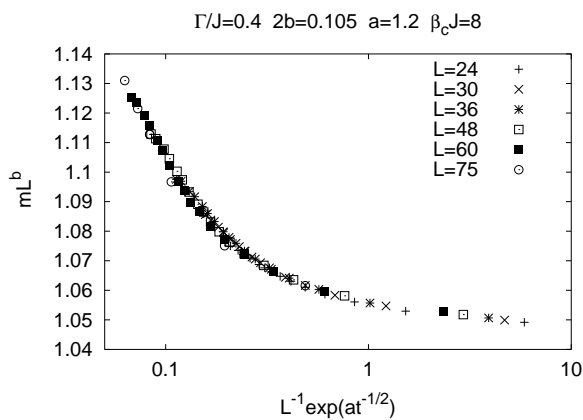}}
\caption{Data collapse of the order parameter $m$ for the lower 
transition.}
\label{fig:lt_fss}}
\end{figure}

In Fig.~\ref{fig:lt_fss}, we show the lower temperature scaled data (see 
Eq.~\ref{m_kt}). The data scales quite well (but not perfectly, and over
a narrower range than for the upper transition) 
with the following
set of parameters: $\beta_{1} J=8.0, a=1.2, c=0.105$. 
The error in determining
the critical temperature is larger than in the high temperature case. We can
conclude that the system has a transition between the KT phase and the ordered
phase at $\beta_{1} J=8.0\pm1.0$. The critical exponent $\eta=0.105\pm0.02$ at
the transition point. This value of the critical exponent $\eta$ is 
again close to
the theoretical prediction $1/9$.

\begin{figure}[ht]
{\centerline{\includegraphics[angle=-90, width=3in]{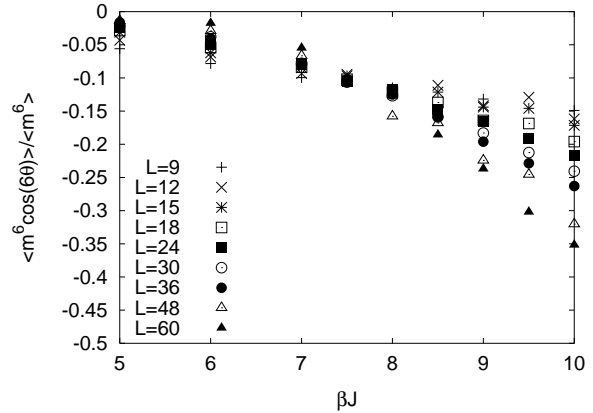}}
\caption{The order parameter $\cos(6\theta)$ versus $\beta J$ for
different system sizes.}
\label{fig:cos}}
\end{figure}

To determine the nature of the ordered phase, we consider
the behaviour of
$\cos(6\theta)$ (Eq.~\ref{cos6theta}). The plot of $\cos(6\theta)$ as
a function of the inverse temperature $\beta$ is shown in
Fig.~\ref{fig:cos}. $\cos(6\theta)$ goes to $-1$ as the temperature
approaches zero. This implies the existence of the  $(+0-)$-phase at low
temperatures for $\Gamma/J=0.4$.

We can also determine the lower transition temperature $T_{1}$ from
the criterion of $\eta =1/9$ at the lower transition point, which
yields $\beta_{1} J=7.5\pm0.5$. At the 
transition temperature thus determined, one finds 
a crossing of $\cos(6\theta)$ as a function of
$\beta$ for different system sizes.\cite{oshikawa} 

\begin{figure}[ht]
{\centerline{\includegraphics[angle=-90, width=3in]{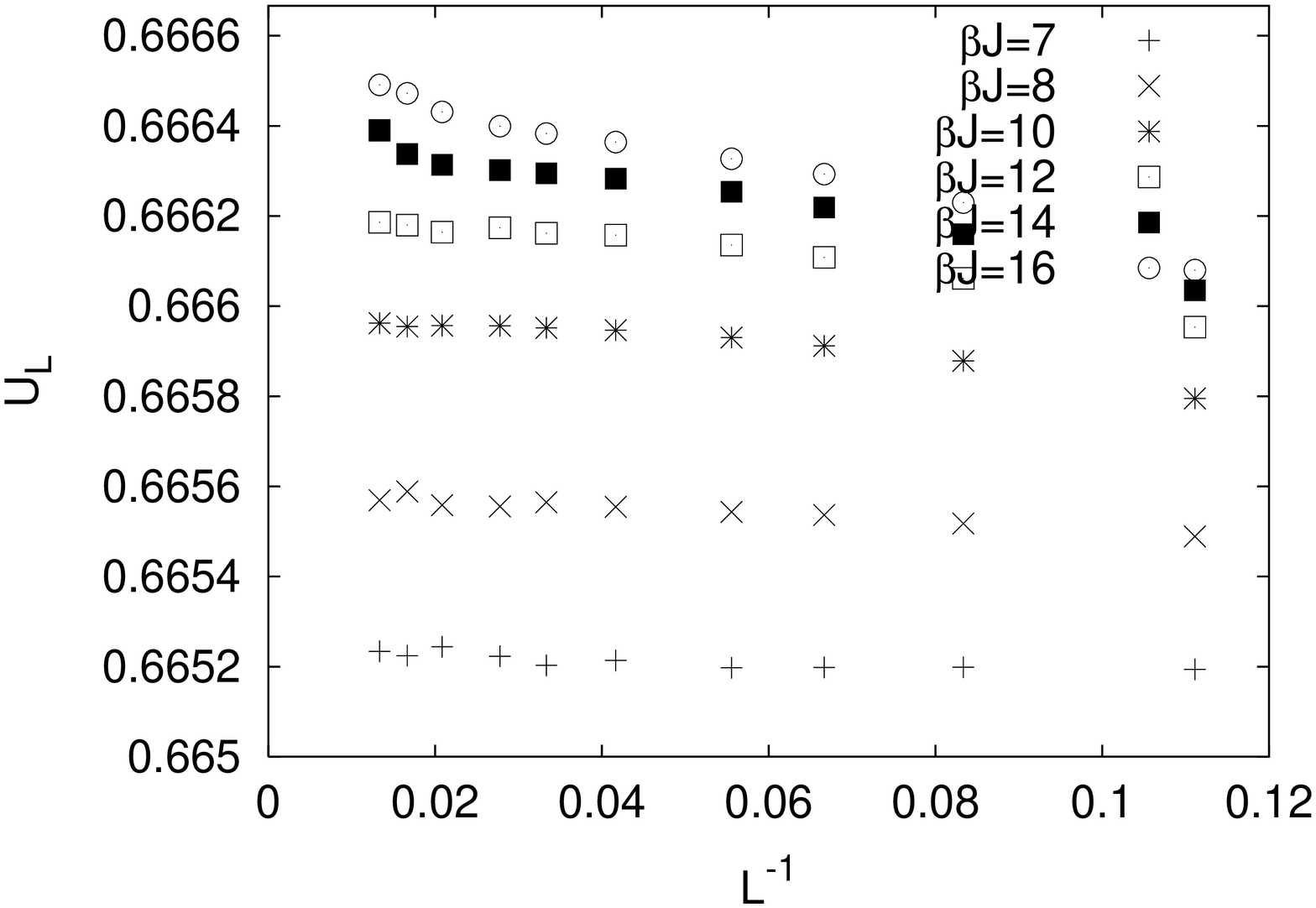}}
\centerline{\includegraphics[angle=-90, width=3in]{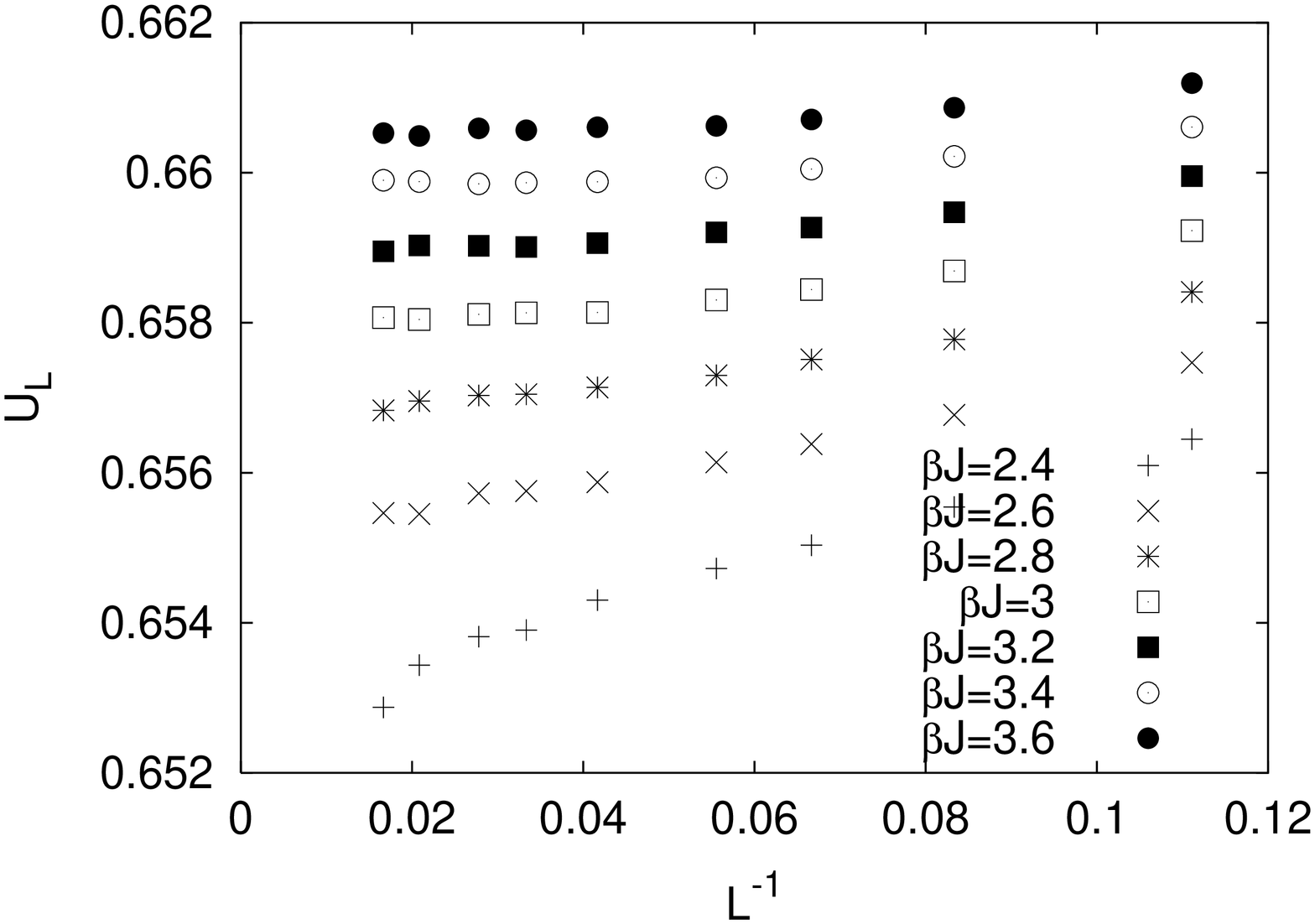}}
\caption{The Binder cumulant $U_L$ versus system size $L$ at various
temperatures. $\Gamma/J=0.4$. Note the different scales for the 
ordinate.}
\label{fig:u}}
\end{figure}

Next, we consider the flow diagram of the Binder cumulant, depicted in
Fig.~\ref{fig:u}. In the high (low) temperature phase, its value
approaches 0 (2/3) as $L\rightarrow\infty$, whereas it levels off to a
value $U^*$ in the KT phase which depends on the strength of the
coupling. Challa and Landau \cite{challalandau} proposed to use this
levelling off as a diagnostic for the KT phase. We do indeed observe
such a levelling over a wide range of temperature at
$\Gamma/J=0.4J$. 

This diagnostic again systematically overestimates the extent of the
critical phase.  This is evidenced by the uppermost curves in
Fig.~\ref{fig:u}. There is an inflection point at large system size
where the value of $U_L$, having apparently levelled off, starts
increasing again. As one approaches the transition from the ordered
phase, this point of inflection wanders to increasingly larger system
sizes and hence beyond the scope of the simulations.

\subsection{Other values of $\Gamma$}

We now describe what we have found for other values of $\Gamma$.  At
small $\Gamma/J\ll 1$, $T/J\ll 1$, $J$ drops out as an energy scale as
the Hilbert space is restricted to the classical ground states, with
excited states frozen out by an energy gap of $O(J)$. The only
dimensionless parameter is then $\beta\Gamma$, so that the lines
emanating from the origin of the phase diagram are straight. We
estimate their slopes to be $T_{KT}^l/\Gamma=0.41\pm0.05$ for the
lower transition, and for the upper:
$T_{KT}^u/\Gamma=0.86\pm0.09$

The higher temperature data collapse does not always give the critical
exponents $\eta$ at critical temperatures close to $1/4$, especially
for large $\Gamma$. For example, we get the following upper set of
critical temperatures and critical exponents $\eta$:
$\beta_{c}J=5.8(6),\eta=0.287(25)$ at $\Gamma/J=0.2$,
$\beta_{c}J=3.5(2),\eta=0.263(15)$ at $\Gamma/J=0.4$,
$\beta_{c}J=2.6(2),\eta=0.230(20)$ at $\Gamma/J=0.8$, and
$\beta_{c}J=3.0(3),\eta=0.203(25)$ at $\Gamma/J=1.2$.  For
$\Gamma/J=0.2$, the high-temperature data scales very well over a wide
range of temperatures. Thus it is difficult to determine the precise value
of the upper transition temperature at this $\Gamma$. 

The systematic trend of decreasing $\eta$ with increasing $\Gamma$ is in
accordance with the fact that $\eta$ at the zero temperature
transition at large $\Gamma$ is different. For the $d=3$ XY
universality class, $\eta$ is in fact close to 0,\cite{zinnjustin} so
that the increasing proximity of this fixpoint should be expected to
show up in a correction of this kind. One can account for the
systematic trend of increasing $\eta$ with decreasing $\Gamma$ in the
same way. Indeed, the frustrated triangular Ising model has a critical
point at $T=0$, where $\eta=1/2$.\cite{stephenson}

\begin{figure}[ht]
{\centerline{\includegraphics[angle=-90, width=3in]{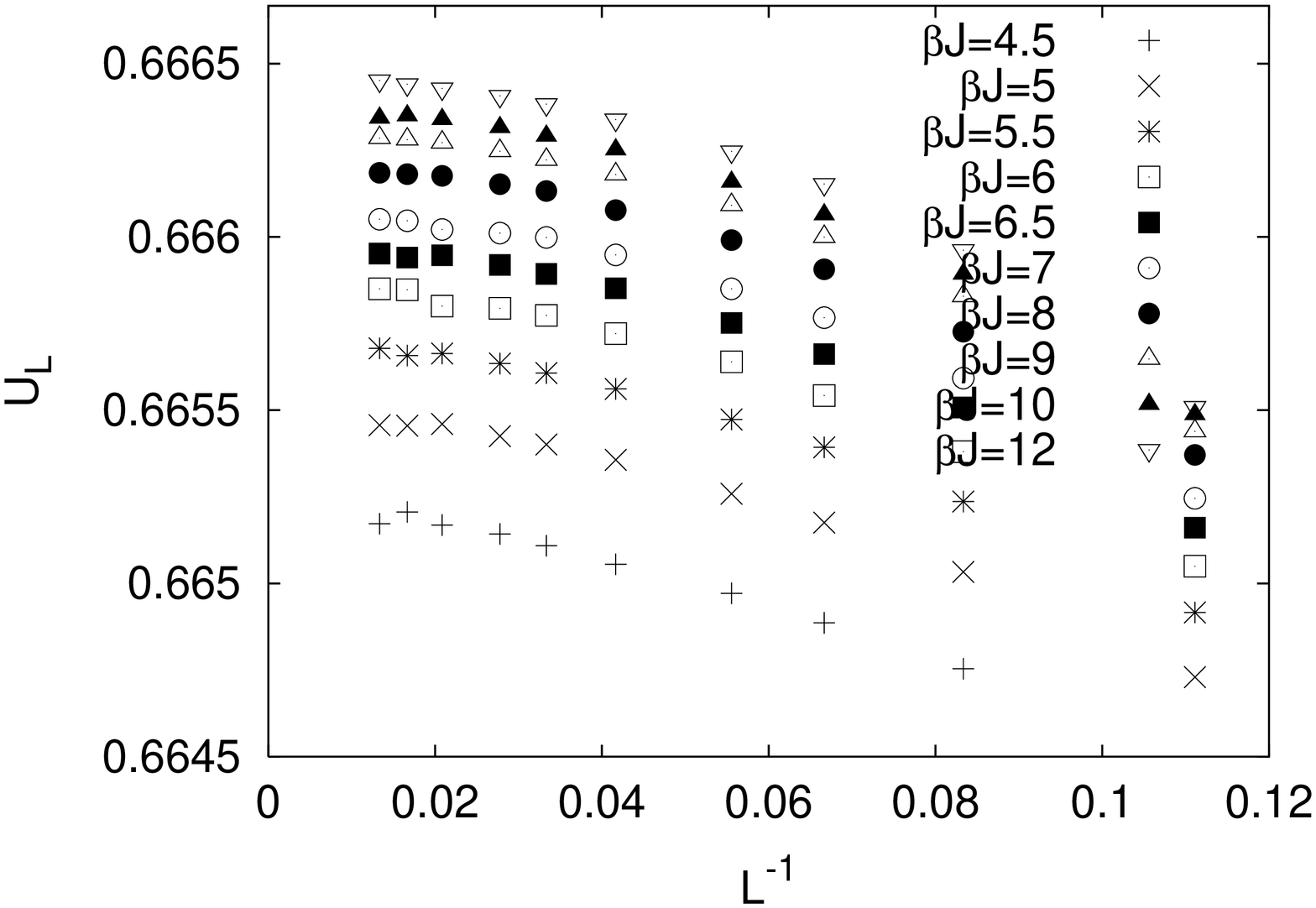}}
\centerline{\includegraphics[angle=-90, width=3in]{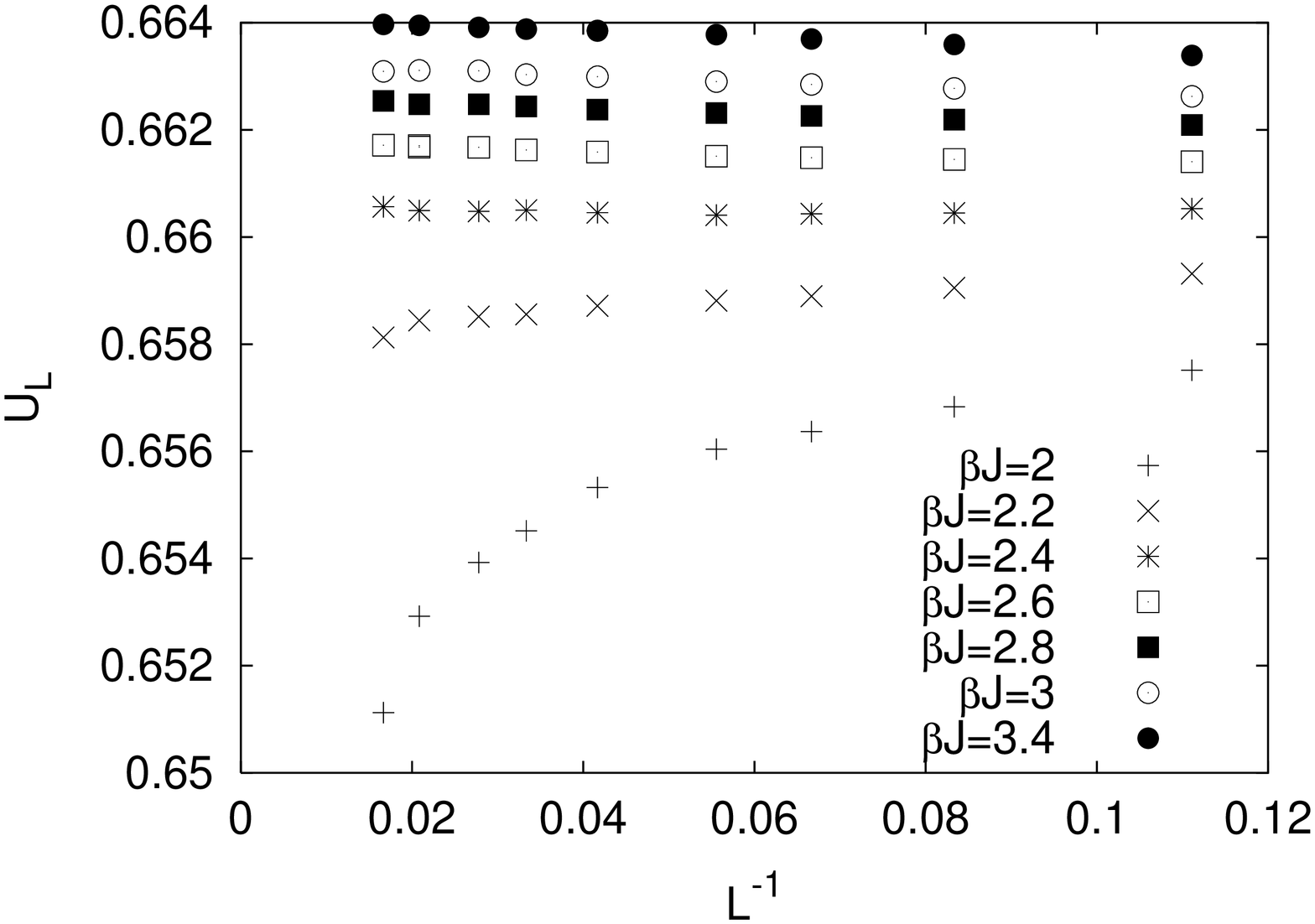}}
\caption{The Binder cumulant $U_L$ versus system size $L$ at various
temperatures. $\Gamma/J=0.8$. Note the different scales for the 
ordinate.}
\label{fig:u2}}
\end{figure}

Probably related to this crossover is the fact that, as $\Gamma$
increases, the plots of the Binder cumulant as a function of system
size fail to display the clear flattening visible in Fig.~\ref{fig:u}
for $\Gamma/J=0.4$ until a system size which is substantially larger;
this is displayed for $\Gamma/J=0.8$ in Fig.~\ref{fig:u2}.

The lower transition at larger $\Gamma$ poses a problem for the
alogrithm we use. Since we used linked lists to store the location of
the domain walls in the temporal direction, the algorithm slows down
considerably as $\Gamma$ increases. Together with the increasingly
slowly decaying correlations in the space direction as the temperature
is lowered, this leads to a considerable increase in the correlation
times of our algorithm.  As a result, the data we were able to obtain
were too noisy to permit a useful data collapse. We can, nonetheless,
try to identify the location of the transition by tracking where $\eta$
crosses 1/9 (Fig.~\ref{fig:phase_diagram})

We note that the classical six-state clock model, which has a phase
diagram very similar to the one we find here, has been studied by many
authors.\cite{challalandau,clockmodel} As is the case for us, almost
all the authors report that it is much more difficult to get data
collapse at the lower KT transition than at the higher.  Only in a
recent paper, Tomita and Okabe,\cite{tomitaokabe} using a
`probability-changing cluster algorithm', have found the lower
transition easily. Given the frustrated and higher dimensional nature
of our spin model, such a cluster algorithm is not available to us.

The nature of the ordered phase at low $\Gamma$ is of the $(+0-)$
type. There is in principle the possibility of a change of sign of
$v_6$ for entropic reasons as the couplings are
varied.\cite{blankschtein,miyashita} As $\Gamma$ increases, it becomes
increasingly hard to determine the sign of $\cos(6\theta)$ for the
system sizes available to us. The largest $\Gamma$ for which we can
confidently state that the lower KT transition is into the $(+0-)$
phase is $\Gamma/J=1.2$. Although this difficulty may
be in part due to a decrease in the strength of $v_6$, there is no
supporting numerical evidence that the sign of $v_6$ ever changes.

\subsection{The quantum critical point}
We now turn to the properties of the quantum critical point, i.e.\ to
the zero temperature transition from the clock symmetry broken ordered
phase to the paramagnet. We can determine its approximate location to
be at $\Gamma_c/J=1.65\pm 0.05$. For large $\Gamma$, the magnet is in
a quantum paramagnetic phase with a size-independent
susceptibility. 

We have not directly attempted to determine the properties of this
transition as extensive previous simulations on the classical stacked
magnet exist. At the time, there were some suggestions that the
observed critical exponents were in fact not those of a $d=3$ $XY$
model\cite{bunker} and perhaps altered due to the presence of another
instability.\cite{heino89,tricrit}

The structure of our phase diagram lends support to the $d=3$ $XY$
universality class scenario via an independent route. By inducing a
dimensional crossover through switching on a finite temperature, we
find that the highly nontrivial phase diagram is that predicted by the
same Landau theory which gives the $d=3$ $XY$ universality class. This
diagnostic is perhaps more robust than a direct determination of the
critical exponents, which can be influenced by corrections to scaling
or the proximity of other instabilities.

The shape of the phase boundary near the quantum critical point
follows from the knowledge of the critical exponents of the quantum
phase transition.\cite{subirqptrev} The boundaries of the KT phase
near $\Gamma_c$ follow the trajectory
$T_{KT}\propto|\Gamma-\Gamma_c|^{\nu z}$, where, for the present case, 
the dynamical critical exponent $z=1$ and $\nu$ is
close to 2/3.\cite{zinnjustin}

\section{Conclusion}
We have demonstrated that the common action of thermal and quantum
fluctuations in the triangular lattice transverse field Ising model
generate an interesting fluctuation-driven phase diagram
including an extended critical phase bordered by a pair of KT
transitions.  We have employed several diagnostics for the presence of
the KT phase and our results are consistent in considerable detail
with what one would expect from an analysis based on a $d=2+1$
dimensional $XY$ symmetric action with a six-fold clock anisotropy for
a temporal dimension of tunable extent.  This provides indirect
support for the $XY$ nature of the quantum critical point. The
dynamical properties of the critical phase in close proximity to the
quantum critical point are a subject worthy of a study in their own
right.

The agreement between numerics and theory, however, is not
complete. There are some deviations from the predicted critical
exponents, which are probably due to a crossover to
the zero temperature critical points. As we
have used several diagnostics to determine the nature and location of
phases and transitions, we are in a position to compare their relative
reliability and find that, for the phase transitions in particular,
considering the system size independence of the Binder cumulant and
the power-law dependence of the magnetisation on system size
systematically overestimate the extent of the critical phase.

In summary, the results we have obtained demonstrate once again that
frustrated magnets provide a good starting point for finding
unconventional phases and phase diagrams. In this particular case, by
using a tunable combination of thermal and quantum fluctuations, we
have managed to realise a standard model from statistical mechanics,
the $XY$ model with sixfold clock anisotropy,\cite{Jose77} in
dimensions $d=2$ and $d=3$, in terms of another one, namely the Ising
model on the triangular lattice. Given other magnets in this class
realise unusual order parameters,\cite{ftfim} this approach should
provide more opportunities for studying exotic phase diagrams based on
simple model spin systems.

\section*{Acknowledgements}
We would like to thank D. Huse, A. Karlhede and especially S. Sondhi
for many useful suggestions and discussions throughout the course of
this work. RM is also grateful to P. Chandra and S. Sondhi for
collaboration on closely related work.  In addition, we thank
A. Bunker, S. Fujiki, T. H. Hansson, J. Lidmar, and M. Oshikawa for
useful discussions. This work was in part supported by the Minist\`ere
de la Recherche et des Nouvelles Technologies with an ACI grant.

\end{document}